\documentclass[]{aa} 
\usepackage{graphicx}
\usepackage{natbib}
\def\hbeta{\mbox{H$\beta$}}
\def\lya{\mbox{Ly$\alpha$}}

\def\ecs{\mbox{~erg~cm$^{-2}$~s$^{-1}$}}

\def\lesssim{\mathrel{\hbox{\rlap{\hbox{\lower4pt\hbox{$\sim$}}}\hbox{$<$}}}}
\def\gtrsim{\mathrel{\hbox{\rlap{\hbox{\lower4pt\hbox{$\sim$}}}\hbox{$>$}}}}

\usepackage{txfonts}
\begin{document}
\title{A jet-cloud interaction in the 3C~196 environment \thanks{Based
    on observations made with the WHT operated on the island of La
    Palma by the Isaac Newton Group in the Spanish Observatorio del
    Roque de los Muchachos of the Instituto de Astrofisica de
    Canarias},\thanks{Based on observations obtained at the
    German-Spanish Astronomical Center, Calar Alto, operated by the
    Max-Planck-Institut f\"ur Astronomie Heidelberg jointly with the
    Spanish National Commission for Astronomy.}}

%\titlerunning{}
   \author{L.~Christensen          \inst{1,2}
          \and K.~Jahnke           \inst{1,3}
          \and L.~Wisotzki         \inst{1}
          \and S.~F. S\'anchez     \inst{1,4}
          \and K.~Exter            \inst{5}
          \and M.~M.~Roth          \inst{1}
          }
   \offprints{L. Christensen}
   \institute{Astrophysikalisches Institut Potsdam, An der Sternwarte 16,
     14482 Potsdam, Germany\\
     \email{lichrist@eso.org} 
     \and European Southern Observatory, Casilla 19001, Santiago 19, Chile
    \and Max Planck Institut f\"ur Astronomie, K\"onigstuhl 17, 69117
Heidelberg, Germany
    \and Centro Astronomico Hispano Aleman de Calar Alto, Spain
    \and Instituto de Astrof\'isica de Canarias, La Laguna, Tenerife, Spain
}

  \date{Received 24 November 2005 / Accepted 7 March 2006}
   
%%%%%%%%%%%%%%%%%%%%%%%%%%%%%%%%%%%%%%%%%%%%%%%%%%%%%%%%%%%%%%%%%%%%   
 \abstract{
Powerful radio galaxies and radio-loud quasars at high redshifts are
frequently associated with extended emission-line regions
(EELRs). Here we investigate the [\ion{O}{ii}] EELR around the quasar
3C~196 at $z=0.871$ using integral field spectroscopy. We also detect
extended [\ion{Ne}{ii}] emission at a distance of about 30 kpc from
the core. The emission is aligned with the radio hot spots and shows a
redshifted and a blueshifted component with a velocity difference of
$\sim$800~km~s$^{-1}$.  The alignment effect and large velocities
support the hypothesis that the EELR is caused by a jet-cloud
interaction, which is furthermore indicated by the presence of a
pronounced bend in the radio emission at the location of the radio hot
spots.  We also report observations of two other systems which do not
show as clear indications of interactions. We find a weaker alignment
of an [\ion{O}{ii}] EELR from the $z=0.927$ quasar 3C~336, while no
EELR is found around the core-dominated quasar OI~363 at $z=0.63$.
     
      }

   \maketitle

%%%%%%%%%%%%%%%%%%%%%%%%%%%%%%%%%%%%%%%%%%%%%%%%%%%%%%%%%%%%%%%%%%%%%%%%%%%%%

\section{Introduction}

 Interactions of powerful radio emission with surrounding gas can
 explain the presence of extended optical emission line regions around
 high redshift, radio-loud AGN.  However, most studies have focused on
 radio galaxies \citep[e.g.][]{vilarmartin97}, rather than radio-loud
 quasars (RLQs), because there is no bright optical glare from the
 central AGN itself. In the case of broad absorption-line quasars,
 measurements of the blue shifted absorption lines have shown large outflow
 velocities \citep{turnshek84}, but only a few observations have
 revealed a direct connection between the more extended gas and the
 central engine. Outflows driven by AGN activity are potentially an
 effective method for quenching star formation in the host galaxy
 \citep{dimatteo05}.

Luminous extended narrow emission-line regions (EELRs), reaching more
than 100 kpc from the QSO nucleus, have been detected around quasars at
redshifts between 1 and 4
\citep[e.g.][]{heckman91b,bremer92b,bremer92}.  At low redshift, 
detailed analyses of the host galaxies are enhanced by observations of
the stellar continuum emission, and in several cases the line emission
extends much farther than the continuum emission. Aside from
interaction with the radio emission, possible explanations include
photoionisation from the central source, or ionising radiation from
massive stars in star-forming regions in the host galaxy.  The
interaction scenario is supported by observations of alignment of the
radio jets with the optical continuum and extended emission line
regions in a sample of radio galaxies \citep{mccarthy87}.  Such
interactions result in shocks which compress and heat the
gas, and subsequent recombination line emission will cool the
material. As alternative explanations for EELRs, remnants from galaxy
mergers \citep{stockton87}, material falling into dark matter
potentials \citep{haiman01}, or cooling flows in massive galaxy
clusters \citep[e.g.][]{fabian90} have been suggested. {

These scenarios make different predictions for the surrounding
material, and so the morphologies and kinematics of the EELRs can be used
to discriminate between them.  To investigate the
structure and kinematics of these EELRs, narrow- and broad-band
images, and slit spectroscopy have traditionally been used.  Integral
field spectroscopy (IFS) presents an alternative technique that allows
imaging and spectroscopy simultaneously.  IFS of six RLQs at redshift
$0.26<z<0.60$ showed that EELRs are common, although not always
aligned with the radio axis \citep{crawford00}.
An alignment between the radio axis and the extended emission around
one lobe-dominated quasar was reported \citep{bremer97}, and later IFS
gave evidence for a jet-cloud interaction \citep{crawford97}.  At
$z>2$, detections of extended [\ion{O}{ii}] and [\ion{O}{iii}]
emission lines have indicated no strong evolution with redshift, and a
tendency for stronger line emission to be spatially coincident with
the stronger radio emission \citep{wilman00}.

The AGN unification scheme states that radio galaxies and RLQs are the
same objects viewed at different angles relative to their jets
\citep{barthel89}.  IFS of radio galaxies have also indicated
jet-cloud interactions through the alignment effect and kinematics
\citep{marquez00,solorzano03,sanchez04b}.  Like the EELRs from RLQs,
the alignment is roughly consistent with the radio morphology
\citep{mccarthy87}. This alignment effect for RLQs and the EELRs is
also found at higher redshifts, but it is not as well-determined at
$z>2$ \citep{heckman91b,hutchings92}, possibly due to resonance
scattering of \lya\ photons.

This paper presents observations of three RLQs at $0.6<z<0.9$ and the
first analysis of the alignment effect of the EELRs in this redshift
range using IFS.

%%%%%%%%%%%%%%%%%%%%%%%%%%%%%%%%%%%%%%%%%%%%%%%%%%%%%%%%%%%%%%%%%%%%%%
%%%%%%%%%%%%%%%%%%%%%%%%%%%%%%%%%%%%%%%%%%%%%%%%%%%%%%%%%%%%%%%%%%%%%%
\section{Observations and data reduction}
\label{sect:3c_obs_data}

The data were obtained in connection with a project aimed at detecting
emission from galaxies associated with intervening damped
Lyman-$\alpha$ absorption lines \citep{christensen04a,christensen04b}.
From the original sample of seven QSOs at $z<2$, only three RLQs have
redshifted strong optical emission lines within the wavelength range
of the observations.

The observations were carried out with two integral field
spectrographs; INTEGRAL \citep{arribas98} mounted on the 4m William
Herschel Telescope, La Palma, and the Potsdam Multi Aperture
Spectrophotometer (PMAS) mounted on the 3.5m telescope at Calar Alto
\citep{pmas00,roth05}.  Table~\ref{tab:3c_log} presents a log of the
observations.

\begin{table*}
\begin{footnotesize}
\centering 
\begin{tabular}{lllllll}
 \hline \hline
   \noalign{\smallskip}
Name  &  alias & redshift &date & instrument & exposure time (s) & seeing (arcsec)\\
  \noalign{\smallskip}
   \hline
   \noalign{\smallskip}
\object{OI 363} & \object{Q0738+313} &0.63 & 2003-04-27 & PMAS   & 2$\times$1800 & 0.8 \\
          &        &      & 2003-04-30 & PMAS   & 2$\times$1800 & 1.2\\
          &        &      & 2004-01-16&INTEGRAL & 4$\times$1800 & 1.0--1.2\\
\object{3C 196} & \object{Q0809+4822}&0.871& 2004-01-16 &INTEGRAL& 8$\times$1800 & 1.0--1.5\\
\object{3C 336} & \object{Q1622+236} &0.927& 2003-04-27 & PMAS   & 6$\times$1800 & 1.0 \\
   \noalign{\smallskip}
\hline
\end{tabular}
\caption[]{Log of the observations. The observations for 3C~196 were
obtained in non-photometric conditions and all derived fluxes are
relative.  }
  \label{tab:3c_log}
\end{footnotesize}
\end{table*}

\subsection{INTEGRAL observations}

The observations were obtained with the SB2 fibre bundle, which
consists of 189 object fibres plus 30 sky fibres arranged in a
ring with a diameter of 90\arcsec.  Each fibre has a diameter of
0\farcs9 on the sky, giving a main field of view of
12\arcsec$\times$16\arcsec\ with a non-contiguous sampling  and a
filling factor of about 67\%. We used a 600 lines mm$^{-1}$ grating
with a dispersion of 3 {\AA} pixel$^{-1}$ and a spectral resolution of
6~{\AA} measured from the width of sky emission lines.

Data reduction was performed using IRAF tasks modified specifically
for the reduction of INTEGRAL data \citep[see][]{garcia05}.  Bias
frames were subtracted and the frames were cleaned for cosmic ray hits
using the algorithm described in \citet{pych03}. All spectra were
extracted using the trace of the 219 spectra on the CCD found from an
exposure of a continuum lamp obtained at the beginning of the night.
Wavelength calibration was done using spectra of emission line lamps
also extracted for each fibre.  The RMS deviation of the wavelength
for sky emission lines for each spectrum was determined to be $<$0.2
{\AA}.  Differences between the individual fibre transmissions as a
function of wavelength were corrected for by modeling the transmission
of sky flat frames obtained at twilight.  Two regions on the CCD were
affected by scattered light but there only in the blue end of the
spectral range and only affecting some of the spectra. Otherwise,
scattered light was found to be negligible compared to the overall
read noise of the CCD.  Some sky fibres are contaminated by the QSO
flux because they are located next to the QSO spectra on the CCD, and
therefore affected by cross talk. The sky fibres were examined, and
those uncontaminated by the QSO flux were averaged and subtracted from
each object spectrum.  We do not expect that any extended line
emission region is present at a distance of 45\arcsec.  To facilitate
inspection and visualisation, the data were interpolated onto a cube
of square {\sl spaxels} (spatial elements) with sizes of
0\farcs3$\times$0\farcs3.  This interpolation and further
visualization of the data cubes was done with the Euro3D visualization
tool \citep{sanchez04a}.  When one-dimensional spectra are extracted,
we retain the original spaxels sizes, instead using the interpolated
data cube.

\subsection{PMAS observations}
PMAS has 16$\times$16 fibres coupled to a lens array and has a
contiguous sampling of the sky.  The effective size of each lens
is square with a size of 0\farcs5 on a side. The instrument
configuration used provided a field of view of
8\arcsec$\times$8\arcsec\ and using a grating with 300 lines mm$^{-1}$
resulted in a spectral resolution of 6.6~{\AA}.

The method for reducing PMAS data was essentially the same as for
INTEGRAL data and was done with our own IDL-based software package P3d
\citep{becker02}. The main difference was the sky subtraction because
PMAS does not have allocated sky fibres. Instead, an average sky
spectrum was created from spectra at the edge of the field of view
uncontaminated by the QSO or the nebulae. Because the spatial position
of the nebulae was previously unknown, several different selections of
sky spectra were examined before selecting an appropriate sky
background spectrum. For the final sky background spectrum we selected
spaxels where no emission lines were detected visually around the
wavelength range of interest,  but we did not detect line emission
for the fibres at the edge of the field. Hence, an over-subtraction
of the extended emission should be a small effect.

Both data sets were flux calibrated by comparison with observations of
spectrophotometric standard stars observed at the same nights and with the
same setup as used for the objects.

\section{Results}
This section presents the results obtained for each of the RLQs. For
reference, the one-dimensional spectra of each QSO created from the
data cubes are shown in Fig.~\ref{fig:3c_qso_spec}.

\begin{figure}
\centering 
\resizebox{\hsize}{!}{\includegraphics[]{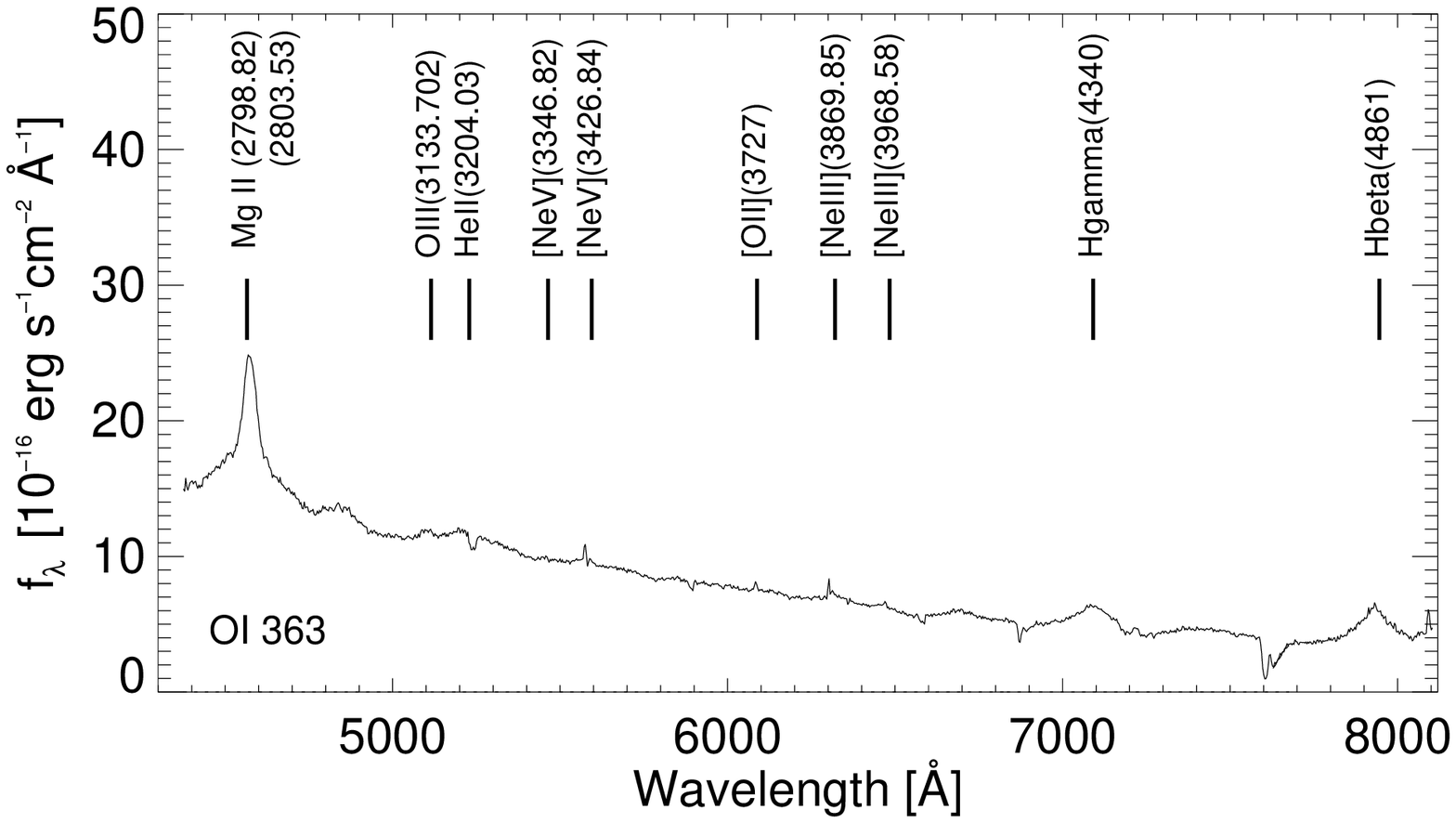}}
\resizebox{\hsize}{!}{\includegraphics[]{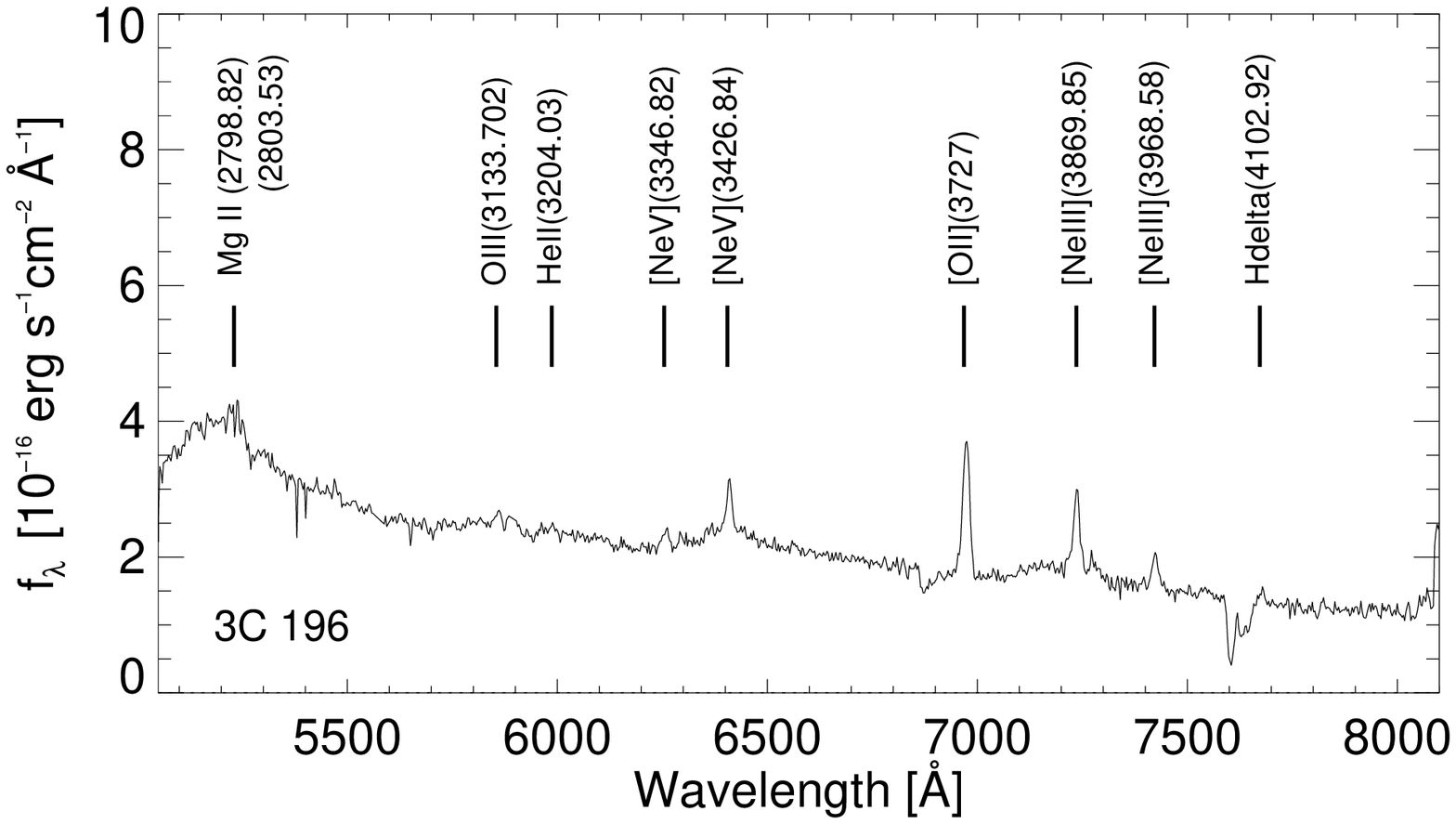}}
\resizebox{\hsize}{!}{\includegraphics[]{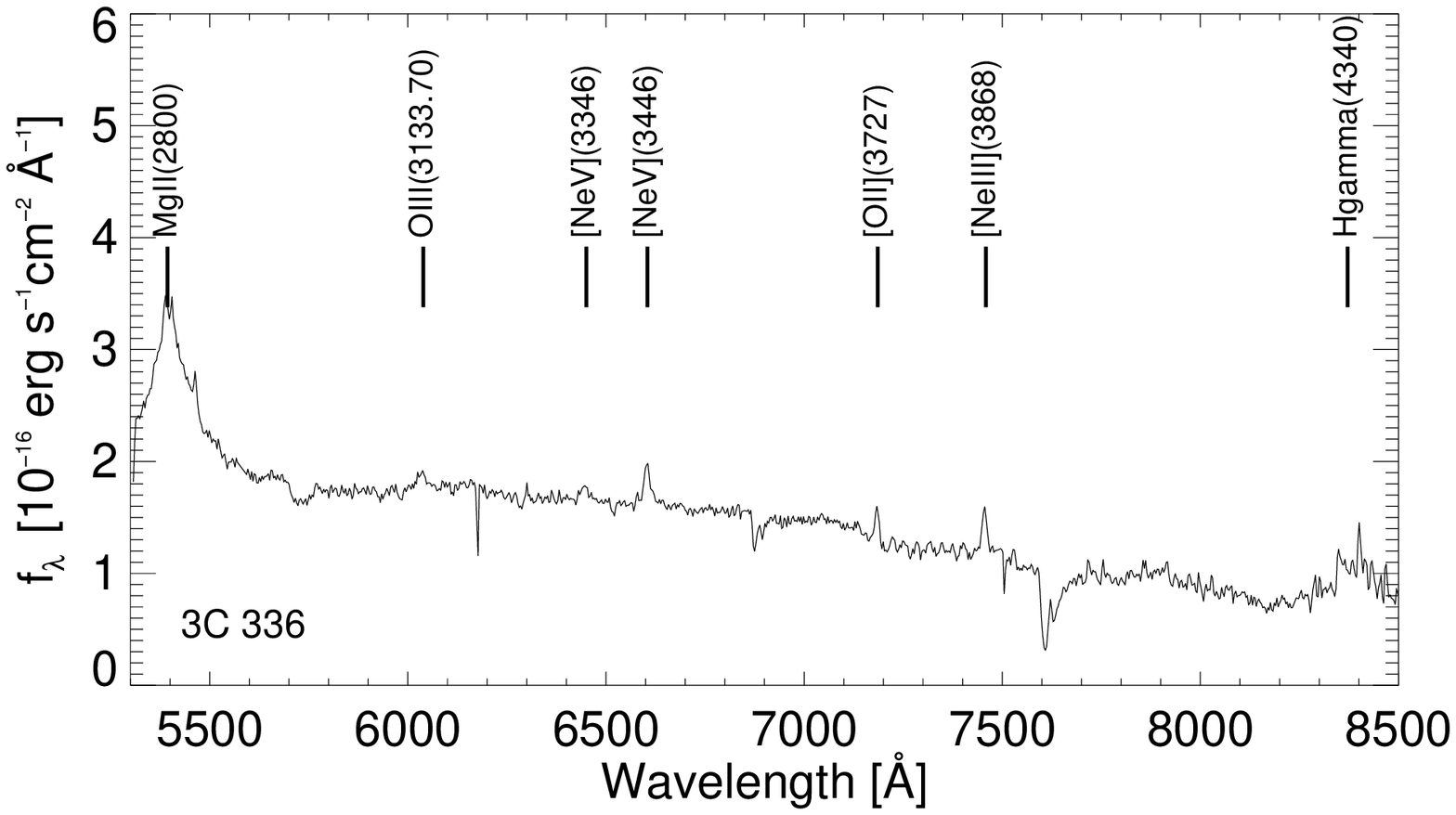}}
\caption{One-dimensional spectra of the quasars extracted from the
  data cubes, by co-adding spectra from 20-30 spaxels. Emission lines
  have been indicated.}
\label{fig:3c_qso_spec}
\end{figure}

To detect the EELRs it is first necessary to subtract the nuclear QSO
emission because this is by far the dominant contribution. To get a
clean spectrum of the EELR emission we use the approach described in
\citet{sanchez04b} and \citep{sanchez06}.  First, a two-dimensional
model of the QSO PSF is made for each slice in the data cube with a
one-pixel width using the program GALFIT \citep{peng02}.  We allow
both a point source representing the continuum (Gaussian component),
and an extended host galaxy component to be present, but find that the
host component is not detected in any of the objects.  Assuming that
the spatial location of the QSO emission varies smoothly with
wavelength we make a model PSF in the form of a data cube. This model
data cube is subtracted from the original date to create a residual
data cube.

\subsection{3C~196}
This is a steep-spectrum, lobe-dominated RLQ \citep{pooley74}, with a
pronounced bend of the radio emission at the location of the hot spots
\citep{brown86}.  Extended oxygen emission has previously been
reported based on long-slit spectroscopic observations
\citep{fabian88}, and the [\ion{O}{ii}] and [\ion{O}{iii}] emission
ratio was interpreted as a cooling flow \citep{crawford89}.
Optical images in the continuum and a 28~{\AA} wide narrow-band
[\ion{O}{ii}] image showed extensions predominantly to the north-west
and south-east \citep{ridgway97}. The extension to the south-east is
mainly caused by an intervening galaxy at $z=0.4367$ which is
responsible for a strong absorption line system in spectrum of the
background QSO \citep{boisse90}.  

The INTEGRAL data cube of this object clearly shows extended
[\ion{O}{ii}] emission, even in the frame where the QSO emission is
not subtracted.  After subtracting the QSO emission, residuals from
the QSO are still present due to uncertainties in the wavelength
dependent PSF determination, but this effect is strong for only the
central region as shown in Fig.~\ref{fig:3c_slices}.  The two radio
hot-spots found at 5 GHz with a separation of 5\farcs8 at
P.A.~$\approx$~120$^{\circ}$ \citep{pooley74} are indicated by the
``+'' signs. The line emission is predominantly located in two regions
as noted before, and the morphology is similar to the [\ion{O}{ii}]
narrow band image presented in \citet{ridgway97}. The north-western
part of the nebula appears to have a continuum counterpart which
extends to about 2\arcsec\ from the QSO \citep{boisse90,ridgway97},
but the IFS data is not sensitive to the continuum.  North of the
quasar we find that the line emission extends out to 3\arcsec\ (22
kpc), while a brighter region extends 5\arcsec\ (36~kpc) to the
south. Here and throughout we have assumed a flat cosmology with
$H_0=70$ km~s$^{-1}$~Mpc$^{-1}$, $\Omega_{m}=0.3$ and
$\Omega_{\Lambda}$~=~0.7.

Co-adding all the spectra associated with the extended emission after
the QSO emission is subtracted results in an emission line with a
total flux of 1.2$\times10^{-15}$~\ecs\ with a \textit{FWHM} of
650$\pm$180~km~s$^{-1}$. The width [\ion{O}{ii}] does not vary
significantly over the face of the nebula within the uncertainties of
the emission line \textit{FWHM}. However, it is expected that a
jet-cloud interaction would result in an increased velocity dispersion
at the location of the hot spots. With the current data we can not
detect this.  Higher spatial resolution data are needed to determine
if this is the case.

Besides [\ion{O}{ii}] emission, we also detect extended
[\ion{Ne}{iii}] $\lambda$3869 emission associated with the brightest
[\ion{O}{ii}] region to the south of the QSO, as shown in the
extracted one-dimensional spectrum in Fig.~\ref{fig:3c_onedspec}.
This is similar to the long-slit spectrum presented in
\citet{steidel97}. The extension of the [\ion{Ne}{iii}] nebula appears
smaller than for the [\ion{O}{ii}] region, but the signal-to-noise
level does not allow for a detailed analysis. The emission line ratio
log([\ion{Ne}{iii}]/[\ion{O}{ii}])~$\approx$~--0.6$\pm$0.1 is larger
than observed for \ion{H}{ii} regions in the Large Magellanic Cloud
\citep{oey00}, where values between --1.6 and --0.7 are found,
indicating that the ionising flux is harder around the RLQ.  Compared
to a few other radio-loud objects \citep{stockton02,solorzano03}, it
shows a similar line ratio.  We note that the two emission line
\textit{FWHM} are very different; 20$\pm$1 {\AA} for [\ion{O}{ii}],
while [\ion{Ne}{iii}] is barely resolved (2.7$\pm$3.3 {\AA} when
corrected for the instrument resolution). This could signify that the
emission originates in different volumes, in which case the line ratio
has no physical meaning.  Alternatively, a jet-cloud interaction could
significantly increase the \textit{FWHM} of the [\ion{O}{ii}] line at
the location of the hot spots, while the [\ion{Ne}{iii}] only arises
from the brighter part of the nebula.

To check for the presence of fainter extended emission farther away
than 5\arcsec\ from the quasar centre, 30 spectra are co-added at
several spatial locations. From the non-detections of emission lines,
we derive an upper limit for emission in the surrounding field of
$\sim$1$\times10^{-17}$~erg~cm$^{-2}$~s$^{-1}$~arcsec$^{-2}$ in the
INTEGRAL data. Not surprisingly, we neither find continuum emission
farther than 5\arcsec\ from the QSO.

\begin{figure}
\centering
\resizebox{6.7cm}{!}{\includegraphics{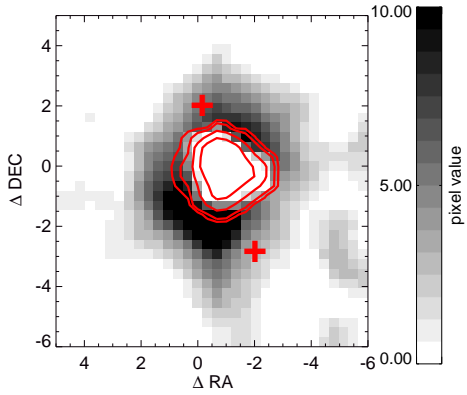}}
\caption{A narrow-band image of the [\ion{O}{ii}] emission at
  $z=0.871$ (6954--6990 {\AA}) from 3C~196 interpolated to a grid
  scale of 0\farcs3. The axes are given in arcseconds, and the gray
  scaling is relative.  Residuals from the QSO subtraction are present
  at the central region, which corresponds to three spaxels.  Contours
  of an image at 7000--7050 {\AA} are overlaid, and the QSO centre is
  located at ($-1,-1$). The innermost contour corresponds to the FWHM
  of the QSO PSF.  The '+' signs denote the positions of the radio
  hot spots.}
\label{fig:3c_slices}
\end{figure}

\begin{figure}
  \centering \resizebox{\hsize}{!}{\includegraphics{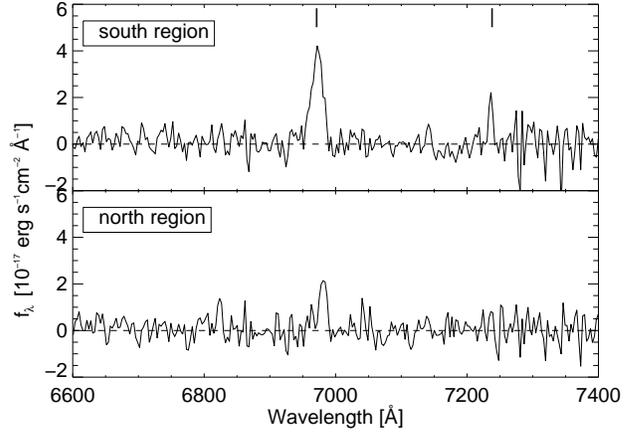}}
\caption{One-dimensional spectrum of the brightest region to the south
  of 3C~196 where both [\ion{O}{ii}] and [\ion{Ne}{iii}] emission
  appear extended.   These lines are indicated by the vertical
  lines. The lower panel shows a one-dimensional spectrum of the
  northern region where no [\ion{Ne}{iii}] emission is detected.}
\label{fig:3c_onedspec}
\end{figure}

%\subsubsection{Velocity structure}
The spectra from various fibres show that the extended emission
around the QSO is progressively shifted in wavelength. 
Fitting Gaussian profiles to the individual spectra gives
evidence for systematic velocity structure over the nebula, as shown in
the upper panel of Fig.~\ref{fig:3c_vel_cont}.  Only spectra with
[\ion{O}{ii}] detections larger than 2$\sigma$ above the background
are shown in colour. Compared to the systemic redshift of the QSO,
the bright southern region has a relative blueshift of $-360$
km~s$^{-1}$, whereas the fainter north western region is redshifted by
up to +500 km~s$^{-1}$. Another representation in the lower panel in
Fig.~\ref{fig:3c_vel_cont} shows a grey scale map of the intensity in
the [\ion{O}{ii}] line emission region derived by fitting emission
lines in individual spectra, and the velocity structures of the
emission are indicated by contours.

\begin{figure}
\centering
\resizebox{7cm}{5.7cm}{\includegraphics[bb=50 350 475 710,clip]{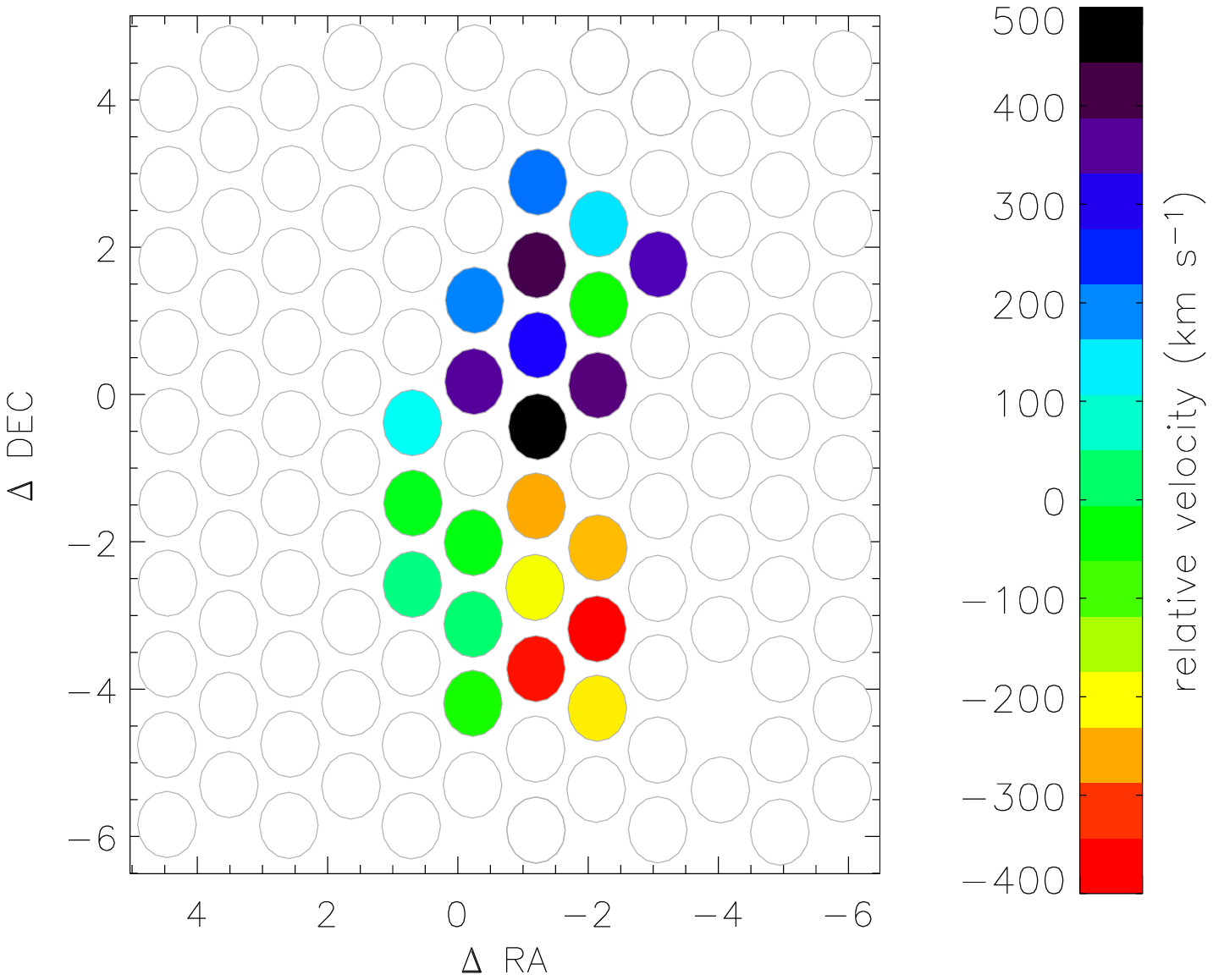}}
\resizebox{7cm}{5.7cm}{\includegraphics[bb=50 350 475 710,clip]{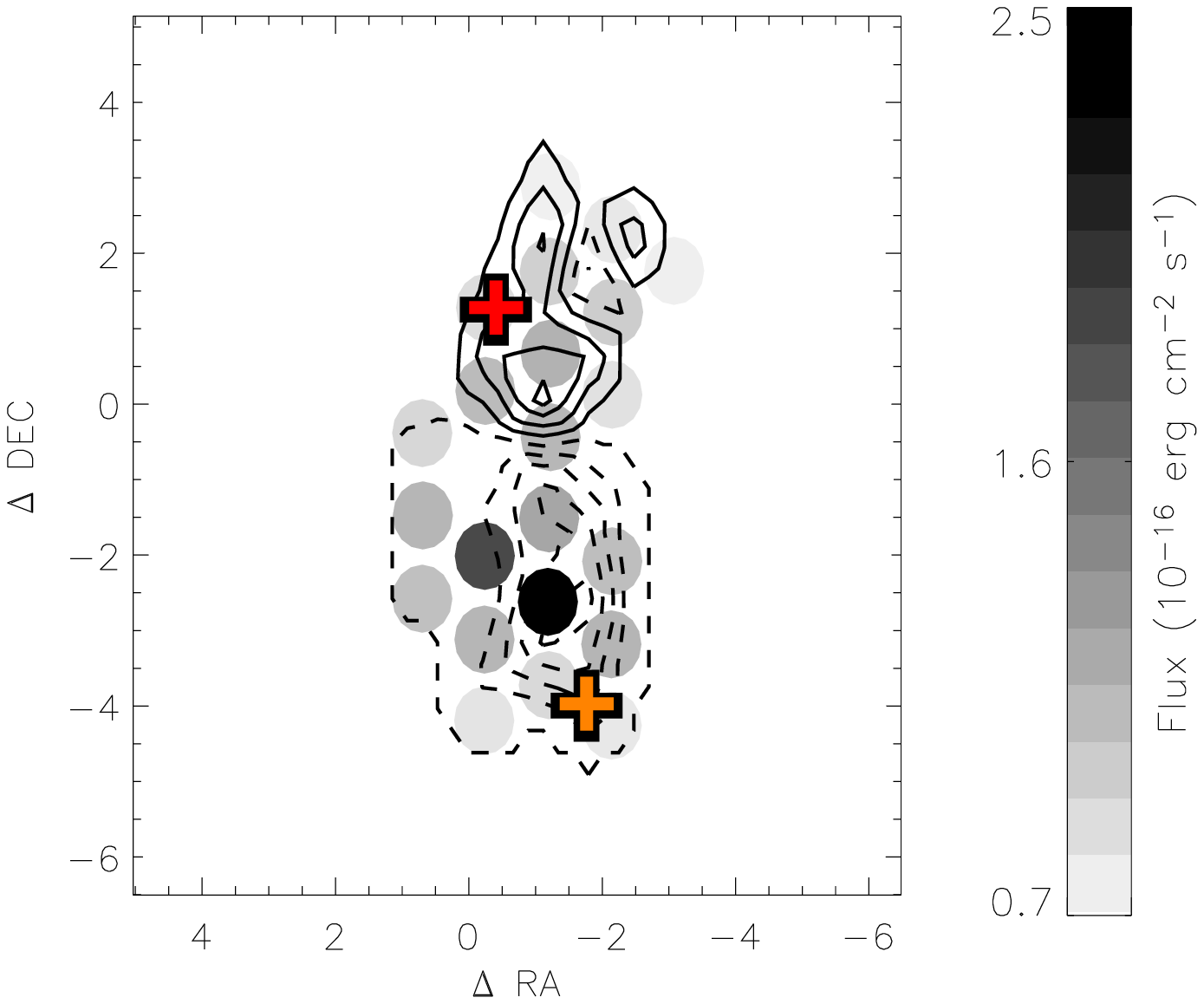}}
\caption{{\sl Upper panel}: Velocity structures of the extended
  [\ion{O}{ii}] emission around 3C~196 determined after the QSO
  emission was subtracted.  The zero-point of the velocity is the
    QSO redshift. Outlined circles indicate the spectra where
  no [\ion{O}{ii}] emission lines could be fit due to their low
  signal-to-noise levels. One spaxel close to the QSO centre has
    not been fitted, because the emission line is affected by QSO
    subtraction residuals. \emph{[See the online edition of the Journal for a colour version of this figure.]}
  {Lower panel}: Image of the [\ion{O}{ii}] line emission intensity
  around 3C~196 with velocity contours overlaid. Contour levels are
  separated by 100 km~s$^{-1}$ and negative velocities are shown by
  the dashed lines. The ``+'' signs indicate the positions of the two
  radio hot-spots which are roughly spatially coincident with the
  brightest emission line regions. }
\label{fig:3c_vel_cont}
\end{figure}

Because our observing campaign was targeted towards finding emission
from a $z=0.4367$ galaxy responsible for the damped \lya\ (DLA), the
wavelength coverage is not optimal compared to similar studies of
other QSOs, which use the [\ion{O}{ii}]/[\ion{O}{iii}] ratio to
derive an internal gas pressure \citep[e.g.][]{crawford00}. Therefore
the present data cube does not allow modeling of the ionising
conditions in the nebulae.

The position of the EELR is not exactly aligned with the hot spots to
the one arcsec accuracy.  However, when we consider that the spectra
do not sample the emission contiguously, both the position of the
northern and the southern radio hot-spots are spatially coincident
with the brightest emission line regions. The southern lobe is
associated with the blueshifted emission, and the northern one with
the redshifted part.  This presents strong indications for an
interaction with the radio jets. However, without polarization
measurements from each hot spot we do not have the information on
which is the nearest, and hence not completely rule out in-flow of the
surrounding gas.

A galaxy responsible for the DLA line at $z=0.4367$ has been detected
1\farcs5 the the south-east of the QSO
\citep{boisse90,lebrun97,chen05}. At this redshift the \hbeta\
emission line falls close to the [\ion{O}{ii}] line at the QSO
redshift, and so potentially the south-western emission could
originate from this DLA galaxy. However, it is unlikely that the
brightest region of emission to the south is contaminated; the most
likely source for the emission is the QSO environment.

\subsection{OI~363}
This is a core-dominated, flat spectrum RLQ \citep{stanghellini97}.
Chandra X-ray observations have revealed an X-ray jet with several
knots extending to the south-east of the quasar
\citep{siemiginowska03}. The end of the X-ray jet at 200~kpc is
spatially coincident with the weaker extended radio emission.

The [\ion{O}{ii}] emission line is within the wavelength range of the
IFS observations, but a subtraction of the QSO emission from the IFS
data cubes reveals no extended emission in either data sets. A
narrow-band image created from the INTEGRAL data cube at the
wavelength region around [\ion{O}{ii}] at $z=0.63$ with a width of
20~{\AA} is shown in Fig.~\ref{fig:3c_0738_map}.  No bright regions
with associated emission lines are found. Other narrow-band images
offset by up to 1000 km~s$^{-1}$ relative to the quasar redshift show
no line emission regions either.  We also note that the nuclear
[\ion{O}{ii}] emission line shown in Fig.~\ref{fig:3c_qso_spec}
appears weak. We estimate an upper detection limit for extended
[\ion{O}{ii}] emission of about $3\times10^{-17}$ \ecs arcsec$^{-1}$
in the INTEGRAL data set and $2\times10^{-17}$ \ecs arcsec$^{-1}$ in
the PMAS data set, based on experiments with artificial emission
lines.

\citet{hutchings92} obtained broad-band and narrow-band images
centered on the [\ion{O}{ii}] emission line at the quasar redshift and
found evidence for a detached region of [\ion{O}{ii}] emission
approximately 2\arcsec\ in size and offset by $\sim$3\arcsec\ to the
north-east. By creating a narrow-band image from the data cube
with the same wavelength range as in \citet{hutchings92} we search for
this detached object, but it is not detected in our data.

\begin{figure}
  \centering 
\resizebox{6cm}{!}{\includegraphics[bb= 48 165 568 560, clip]{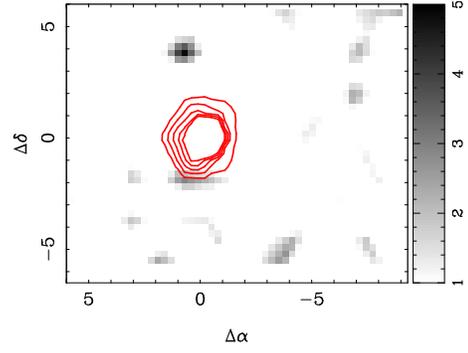}}
\caption{Narrow-band image of OI~363 created from the INTEGRAL data at
  $6070<\lambda<6090${\AA} which includes [\ion{O}{ii}] at
  $z=0.63$. The scaling has been chosen to show features brighter than
  2$\sigma$ above the background noise, and no EELR features
  surrounding the quasar are detected.  The small features visible in
  this image are not significant, i.e. they have no associated
  emission line in an extracted one-dimensional spectrum at this
  wavelength. The contours show an off-band narrow-band image of the
  QSO centered at (0,0) where the innermost contour represents the
  seeing \textit{FWHM}. Because of cross talk between spectra
  positioned next to each other on the CCD, the QSO PSF appears non
  circular.}
\label{fig:3c_0738_map}
\end{figure}

\subsection{3C~336}
3C~336 is a steep-spectrum lobe-dominated RLQ with radio lobes
separated by 28\arcsec\ at PA~$\approx$~30$^{\circ}$ and a jet
extending to the south-east of the QSO \citep{pooley74,bridle94}. Both
\citet{bremer92b} and \citet{steidel97} reported extended
[\ion{O}{ii}] emission at $z=0.927$ to the north and south of the
QSO. This emission, detected out to a distance of 5\arcsec\ from the
core, was previously interpreted as due to a cooling flow
\citep{bremer92b}, while \citet{steidel97} found that it belonged
to a blob without a continuum counterpart.  The quasar is located in
a region where many galaxies are present close to the sight-line.
Most of the galaxies are intervening, but seven galaxies have
redshifts consistent with being associated with a cluster at
$z=0.923$, i.e.  at the quasar redshift \citep{steidel97}.

Fig.~\ref{fig:3c_1622} shows a narrow-band image created from the PMAS
data cube. The morphology of the emission region is similar to the
28~{\AA} wide [\ion{O}{ii}] narrow-band image presented in
\citet{ridgway97}, but like these authors we cannot confirm the
extended emission reaching towards the north.  The extended emission
region appears to the south and north-west of the QSO, with the
brightest region towards the south.  The latter is spatially
coincident with the radio jet, whereas the north-west emission line
region is not aligned.

The PMAS data only allows detection of the emission to $\sim$2\arcsec\
from the QSO, and a detailed analysis of the kinematics of the
emission line region is not possible with the signal-to-noise level in
each individual spectrum from the data cube. In the co-added
one-dimensional spectrum for the whole emission line region the total
line flux measured is $(3.8\pm1.4)\times10^{-16}$ \ecs\ after a
correction for Galactic extinction is applied
\citep{schlegel98}. After correcting for the instrumental resolution,
the line width is 910$\pm$280~km~s$^{-1}$.

A narrow-band image from the data cube, offset to slightly longer
wavelengths (7190--7206 {\AA}) shows that the emission feature extends
further to the north-west and is spatially coincident with the
emission feature denoted `2' in a continuum image of
\citet{steidel97},  which is marked by '+' in
Fig.~\ref{fig:3c_1622}.  A one-dimensional spectrum of that region in
the data cube indicates $z=0.9300\pm0.0009$, in agreement with
$z=0.931$ measured by \citet{steidel97}. Furthermore, the data cube
allows a confirmation of their object `3' at $z=0.892$, 3\arcsec\ to
the east of the QSO. Possibly, the emission to the north-west is
unrelated to the QSO and only belongs to the galaxy `2'; thus we
determine the properties of the southern region only. Here the flux is
$1.8\pm0.5\times10^{-16}$ \ecs\ and the \textit{FWHM} is 720$\pm$200
km~s$^{-1}$.  The redshift measured is $z=0.9274\pm 0.0007$, 
corresponding to a velocity difference of 60$\pm$110 km~s$^{-1}$
relative to the QSO redshift.

\begin{figure}
\centering
\resizebox{6cm}{!}{\includegraphics[]{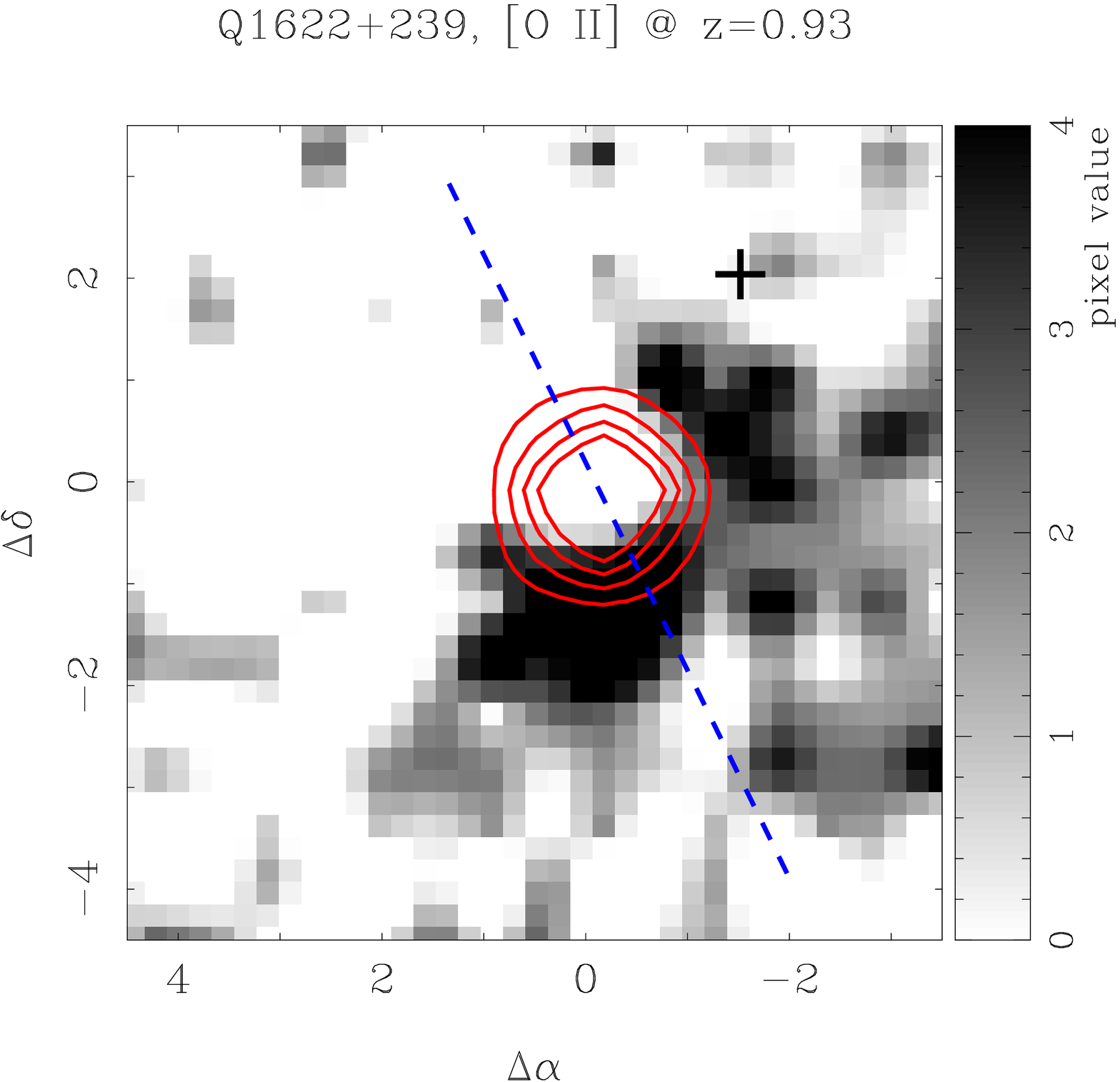}}
\caption{Narrow-band image of the [\ion{O}{ii}] line emission region
  from 3C~336 in the region 7167--7193~{\AA}. The image is
  interpolated to a pixel scale of 0\farcs2\ to make visible the faint
  emission. The field of view is 8\arcsec$\times$8\arcsec, north is up
  and east to the left. The location of the object denoted `2' in
  \citet{steidel97} is marked by the '+' sign.  A narrow-band image
  at slightly redder wavelengths (7190--7206 {\AA}) shows emission at
  larger distances than in this image, which is consistent with the
  2\farcs58 offset reported in \citet{steidel97}. Contours show an
  off-band narrow-band image (6950--7000 {\AA}) where the innermost
  contour represent the seeing \textit{FWHM}. The dashed line
  represents the orientation of the radio jet which extends beyond the
  field of view of the IFS data.  }
\label{fig:3c_1622}
\end{figure}

%%%%%%%%%%%%%%%%%%%%%%%%%%%%%%%%%%%%%%%%%%%%%%%%%%%%%%%%%%%%%%%%%%%%%%
%%%%%%%%%%%%%%%%%%%%%%%%%%%%%%%%%%%%%%%%%%%%%%%%%%%%%%%%%%%%%%%%%%%%%%
\section{Discussion and conclusions}
We have presented integral field spectroscopy of three RLQs, where one
is core-dominated and the other two are lobe-dominated. Extended
[\ion{O}{ii}] emission is seen around the two lobe-dominated RLQs, and
the emission is spatially coincident with the radio emission.  All
lines from the EELRs are relatively narrow with \textit{FWHM}$<$1000
km~s$^{-1}$. The brightest object (3C~196) has an EELR velocity
structure which changes by more than 800 km~s$^{-1}$ over
approximately 60 kpc.  

Compared to the other RLQs where an alignment has been detected
\citep[e.g.]{crawford00}, the EELR around 3C~196 suggests the presence
of a jet-cloud interaction with strong velocity differences.  The
emission line nebula has both a red- and a blueshifted component
aligned with the radio hot spots, and the velocity is larger than
expected for a gravitational origin.  Rotation velocities for the most
rapidly rotating massive spiral disks are around 300 km~s$^{-1}$ at a
distance of about 10 kpc. If the EELR is a rotating disk, its mass
would have to be at least twice that of such a massive spiral.  The
bend of the radio emission at the location of the hot spots noted in
\citet{brown86} supports the interpretation of an interaction. 
The [\ion{O}{ii}] emission from the galaxy is spatially offset from
the intervening galaxy found by \citet{boisse90}, which at a redshift
of 0.87 has \hbeta\ emission lines very close to the [\ion{O}{ii}] at
the QSO redshift. With emission line fitting we are able to separate
the \hbeta\ emission line from the [\ion{O}{ii}] emission, because
there is a 10 {\AA} difference between the two lines, and the
positions are not exactly spatially coincident. 

Extended [\ion{Ne}{iii}] emission is detected for 3C~196 at a distance
of 30~kpc from the quasar nucleus, but only in the southern region
where the [\ion{O}{ii}] emission is strongest.  Because the
[\ion{Ne}{iii}]/[\ion{O}{ii}] line flux ratio is larger than observed
in \ion{H}{ii} regions, harder ionising radiation must be present if
the emission originates in the same volume.  For a detailed modeling
more emission lines need to be observed.  Ionisation conditions can in
principle be analysed in the EELRs provided that the [\ion{O}{iii}]
emission line is detected too.  Combining the observed emission line
ratios with knowledge of the quasar ionising flux allows determination
of the nebular pressure through photoionisation modeling. However,
this approach is only justified when emission lines arise in the same
medium. This is sometimes neglected, as pointed out by
\citet{stockton02}, who find that the [\ion{O}{ii}] emission arises in
a much denser medium than [\ion{O}{iii}] in one RLQ.  
Measurements of other line ratios could help to determine the nature
of the nebulae and the presence of interaction signatures
\citep[e.g.][]{garcia05}.

The alignment of the nebular emission line regions with the radio jets
indicates that some interaction could be present.  To determine
whether the line emission is created by the ionising quasar flux which
is collimated along the radio jet, it is necessary to determine the
underlying continuum flux. If a significant stellar emission component
is present then one can rule out this scenario. No extended continuum
emission is detected in either object analysed here, but if it were
present, the bright nuclear flux complicates the detection. Deeper
(imaging) data could resolve this issue. Furthermore, if the extended
emission is caused by an interaction with the radio jet, larger field
of view observations of 3C~336 are necessary to analyse the effects
at larger distances corresponding to the widely separated radio-lobes.

In the unification scheme, the core-dominated RLQs have their radio
jet orientation closely aligned with the sight line. If radio jets
cause interaction with the surrounding material and give rise to line
emission, a core-dominated RLQ would have less extended emission than
lobe-dominated RLQs.  No EELR is detected for the one core-dominated
RLQ analysed here.

An alternative scenario where the extended emission is dominated by
the QSO ionising radiation directed in a cone
\citep{haiman01,weidinger05} can not be rejected completely on the
basis of this investigation. However, the velocity structure of the
nebulae around 3C~196 supports the hypothesis that it is the
interaction of the material with the radio jet which is the dominant
effect and causes an outflow of the material.

\begin{acknowledgements}
  L.~Christensen acknow\-ledges support by the German Verbundforschung
  associated with the ULTROS project, grant
  no. 05AE2BAA/4. S.F.~S\'anchez and K. Exter acknowledge the support
  from the Euro3D Research Training Network, grant
  no. HPRN-CT2002-00305. K. Jahnke acknowledges support from
  DLR. project no.  50~OR~0404.
\end{acknowledgements}

\bibliography{ms4578}
\end{document}